\documentclass[preprints,article,accept,moreauthors,pdftex,10pt,a4paper]{mdpi}

\def\e{\mathrm{e}}
\def\ii{\mathrm{i}}

\firstpage{1} 
\pubvolume{xx}
\issuenum{1}
\articlenumber{1}
\pubyear{2018}
\copyrightyear{2018}
\history{}

\pdfoutput=1

\Title{
Kick and fix: the roots of quantum control}

\Author{Paolo Facchi $^{1,2}$*\orcidA{} and Saverio Pascazio $^{1,2,3}$\orcidB{}}

\AuthorNames{Paolo Facchi and Saverio Pascazio}

\address{%
$^{1}$ \quad Dipartimento di Fisica and MECENAS - Universit\`{a} di Bari, I-70126 Bari, Italy \\
$^{2}$ \quad INFN, Sezione di Bari, I-70126 Bari, Italy \\
$^{3}$ \quad Istituto Nazionale di Ottica - Consiglio Nazionale delle Ricerche (INO-CNR), I-50125 Firenze, Italy}

\corres{Correspondence: paolo.facchi@ba.infn.it}

\abstract{
When two operators $A$ and $B$ do not commute, the calculation of the exponential operator $\e^{A+B}$ is a difficult and crucial problem. The applications are vast and diversified: to name but a few examples, quantum evolutions, product formulas, quantum control, Zeno effect. The latter are of great interest in quantum applications and quantum technologies. We present here a historical survey of results and techniques, and discuss differences and similarities. We also highlight the link with the strong coupling regime, via the adiabatic theorem, and contend that the "pulsed" and "continuous" formulations differ only in the order by which two limits are taken, and are but two faces of the same coin.}


\begin{document}

\section{Introduction}

Quantum control is a relatively novel discipline, in which one seeks to control the evolution or the final state of a quantum system or to determine the output of a quantum process. The main objective is to counter the detrimental effects of (uncontrollable) noise and environment. The inspiration goes back to classical control, where one would make some measurement of the output and then use the results of such measurement to obtain the desired outcome (feedback mechanism). Similar ideas (but different techniques) motivated the development, in the 1970s and 1980s, of nuclear magnetic resonance imaging, towards very successful applications in medical diagnosis. The objective in that case was to control (spin) coherence. 

In quantum mechanics and quantum applications things become more involved, because a quantum measurement collapses the wave function, thereby strongly affecting the quantum evolution. The question becomes: what can one actually do to a quantum system (from carefully tailoring measurements to accurately designing Hamiltonians or even Lindbladians) that will enable some control of the output? In an even more recent twist, some researchers have also considered the option of  engineering the environment itself, towards specific quantum goals.

The quantum control problem translates to rotating a vector state from a specified initial direction to a desired final direction in a certain time. In a more general setting, this means evolving a generic state (density matrix) from a given initial state to a desired final state in a certain time. This is a difficult task,  which essentially consists in the construction of a suitable transformation, in agreement with the guiding principles of quantum mechanics. We highlight in this article  two main techniques that are studied nowadays.

\section{Divide et impera (Divide and conquer)}

\subsection{History}

In 1948 Feynman, working on his path-integral formulation of quantum mechanics 
\cite{ref:Feynman(1948),ref:FeynmanHibbs(1965),Schulman1981}
noticed that one can obtain the quantum evolution by applying the following formula
\begin{eqnarray}
\label{eq:feynman}
U_N (t)
& = & \underbrace{ 
\left(\e^{-\ii \frac{t}{N} T} \e^{-\ii \frac{t}{N} V}\right)
\left(\e^{-\ii \frac{t}{N}T } \e^{-\ii \frac{t}{N} V}\right)
\cdots 
\left(\e^{-\ii \frac{t}{N} T} \e^{-\ii \frac{t}{N} V}\right)}_{N \; \mathrm{times}} 
\nonumber \\
& = & \left(\e^{-\ii \frac{t}{N} T} \e^{-\ii \frac{t}{N} V}\right)^N 
\to \e^{-\ii t (T+V)}
= \e^{-\ii t H},  \qquad \text{as } N\to\infty,
\end{eqnarray}
where $H=T+V$ is the Hamiltonian of the system, $T$ and $V$ being respectively the kinetic and potential energy. Feynman strived to bypass a crucial difficulty that plagues (exponentiated) quantum operators. Formula~(\ref{eq:feynman}) is valid although $T$ and $V$ do not commute: the key observation is that the error one makes in approximating the evolution with the $N$-product is of order $1/N$, and disappears in taking the limit.

Feynman built on a crucial observation that Dirac had made 15 years earlier~\cite{ref:Dirac1933}, concerning the infinitesimal expression of the quantum propagator. Feynman made no attempt at rigor (beyond the reasonable level of theoretical physics), unlike (presumably) Dirac, who did not dare to go beyond infinitesimal evolutions.

In fact, mathematicians independently worked on what they called product formulae.
Trotter in 1959~\cite{ref:Trotter} and Kato in 1978~\cite{ref:KatoTrotter} investigated the limits of products of the following type
\begin{eqnarray}
\label{eq:trotter}
V_N (t) 
& = & \underbrace{ 
\left(\e^{A/N} \e^{B/N}\right)
\left(\e^{A/N} \e^{B/N}\right) 
\cdots 
\left(\e^{A/N} \e^{B/N}\right)}_{N \; \mathrm{times}} 
\nonumber \\
& = & \left(\e^{A/N} \e^{B/N}\right)^N  
\to \e^{A+B} ,  \qquad \text{as } N\to\infty,
\end{eqnarray}
and proved convergence for certain unbounded (linear) operators $A$ and $B$. 
Both Trotter and Kato were extending results by Lie~\cite{ref:Lie}, who had proved the validity of Eq.~(\ref{eq:trotter}) for $n \times n$ matrices.

\subsection{Zeno}

Misra and Sudarshan's formulation of the quantum Zeno effect~\cite{ref:QZEMisraSudarshan,ref:QZEreview-JPA} makes use of the following formula
\begin{eqnarray}
\label{eq:zenoms}
U_N (t)
& = & \underbrace{ 
\left(\e^{-\ii \frac{t}{N} H} P\right)
\left(\e^{-\ii \frac{t}{N} H} P\right)
\cdots 
\left(\e^{-\ii \frac{t}{N} H} P\right)}_{N \; \mathrm{times}} 
\nonumber \\
& = & \left(\e^{-\ii \frac{t}{N}  H} P\right)^N  
\to \e^{-\ii t H_\textrm{Z}} P,  \qquad \text{as } N\to\infty,
\end{eqnarray}
where $P$ is a projection operator and $H_\textrm{Z}=PHP$ the "Zeno" Hamiltonian~\cite{ref:QZS}.
Notice that in general $U_N (t)$ is not unitary on the range of $P$, while the limit is, under suitable conditions and in particular cases, such as bounded $H$ or finite-dimensional $P$~\cite{ref:artzeno}. A general proof of the above formula is still missing, together with a rigorous definition of the Zeno Hamiltonian for infinite-dimensional projections and unbounded Hamiltonian. 

Interestingly, the above formula can be re-written in terms of an absorbing "optical potential". Let 
$P+Q=1$, with both $P$ and $Q$ projections. Then, by making use of the identity (valid in general, but we take $\gamma >0$)
\begin{equation}
\label{eq:pq}
\e^{-\gamma Q} = P + \e^{-\gamma} Q ,
\end{equation}
one gets
\begin{eqnarray}
\label{eq:zenogamma}
U_N (t)
& = & \underbrace{ 
\left(\e^{-\ii \frac{t}{N} H} \e^{- \gamma Q}\right)
\left(\e^{-\ii \frac{t}{N} H} \e^{- \gamma Q}\right)
\cdots 
\left(\e^{-\ii \frac{t}{N} H} \e^{- \gamma Q}\right)}_{N \; \mathrm{times}} 
\nonumber \\
& = & \left(\e^{-\ii \frac{t}{N} H} \e^{- \gamma Q}\right)^N 
\to  \e^{-\ii t H_\textrm{Z}} P, \qquad \text{as } N\to\infty,
\end{eqnarray}
which is essentially the same as~(\ref{eq:zenoms}). In a few words, the above formula works because the imaginary optical potential $\ii \gamma Q$ quickly (on a timescale $1/N\gamma$) absorbs away the unwanted component of the wave function, acting as a projection $P$.

Observe also that formula~(\ref{eq:zenoms}) is at the basis of the Faddeev-Popov method 
\cite{ref:FaddeevPopov}  to quantize gauge quantum field theories within the framework of the path integral formulation. In this case, the projection imposes the gauge condition.

The afore-mentioned formulas are at the basis of the "pulsed" formulation of quantum control. Physically, the validity of the constraint is preserved during the evolution, but the quantum system is free to move within the (Zeno) subspaces of the (multidimensional constraint)~\cite{ref:QZS}. In applications, $P$ must be designed in such a way as to reduce (or hinder) decoherence.

\subsection{Kicks}
\label{sec-kickform}

In a similar spirit, physically equivalent dynamics can be obtained for a quantum system undergoing repeated unitary "kicks", the physical 
duration of the kick being the shortest timescale of the problem (notice the analogy with a projection, that is also supposed to take
place instantaneously). The evolution reads
\begin{eqnarray}
U_N(t) &=&
\underbrace{\left( \e^{-\ii \frac{t}{N} H} U_{\mathrm{kick}}
\right)\left(\e^{-\ii \frac{t}{N} H} U_{\mathrm{kick}}
\right)\cdots \left(\e^{-\ii \frac{t}{N} H} U_{\mathrm{kick}}
\right)}_{N\; \mathrm{times}} \nonumber \\
& = &
\left(\e^{-\ii \frac{t}{N} H} U_{\mathrm{kick}} \right)^N
\sim \e^{-\ii t H_{\mathrm{Z}}} U_{\mathrm{kick}}^N, 
 \qquad \text{as } N\to\infty,
\label{eq:BBevol}
\end{eqnarray}
where
\begin{equation}
\label{eq:eqHz}
H_{\mathrm{Z}} = \sum_n P_n H P_n
\end{equation}
is the Zeno Hamiltonian and $P_n$ the spectral projections of the kick (taken with a discrete spectrum)
\begin{equation}
U_{\mathrm{kick}}=\sum_n \mathrm{e}^{-\ii\lambda_n} P_n . \quad
(\mathrm{e}^{-\ii\lambda_n}\neq \mathrm{e}^{-\ii\lambda_l}, \; \mbox{for} \; n\neq l.)
\label{eq:specdec}
\end{equation}
This is again a Zeno dynamics. The Zeno subspaces are now a consequence of rapidly oscillating phases between different eigenspaces of the
kick, yielding a superselection rule.
This phenomenon was discovered in chemical physics, where it is known as NMR~\cite{anderson,ernst,freeman,levitt}, and (re)baptized "bang-bang" control in the quantum-information literature~\cite{viola98}. A very nice review is 
\cite{lidarrev}.

 Finally, we mention that the same formulas are of crucial importance in the study of quantum chaos~\cite{ref:CasatiChaos,ref:BerryChaos,ref:Gutzwiller}. 
Set $t= N \tau$, keep $\tau$ fixed, and let $U_{\mathrm{kick}}=\e^{-\ii \tau_0 {V}}$, where $\tau_0$ is  needed for dimensional purposes. The evolution is
\begin{eqnarray}
\label{eq:qchaos}
& & \underbrace{ 
\left(\e^{-\ii \tau T} \e^{-\ii \tau_0 {V}}\right)
\left(\e^{-\ii \tau T} \e^{-\ii \tau_0 {V}}\right)
\cdots 
\left(\e^{-\ii \tau T} \e^{-\ii \tau_0 {V}}\right)}_{N \; \mathrm{times}} 
  =\left(\e^{-\ii \tau T} \e^{-\ii \tau_0 {V}}\right)^N ,  \qquad \text{as } N\to\infty,
\end{eqnarray}
where $T$ is a kinetic energy operator and $V$ a potential. In the above formula $V$ is not divided by $N$, so that the evolution is governed by a singular time-dependent Hamiltonian $H=T + \tau_0  \sum_n \delta (t - n\tau) V$. Notice that unlike in formula (\ref{eq:BBevol}), since $\tau$ is kept fixed, this is a large-time limit $t=N\tau \to \infty$. 
Alternatively, the evolution~(\ref{eq:qchaos}) can be thought as a sequence of purely kinetic and purely potential motions in which alternately the force is switched off (for a time $\tau$) and the mass made infinite (for a time $\tau_0$)~\cite{ref:BerryChaos}.

\section{Persuade et rege (Persuade and rule)}
\label{sec-contform}

Short timescales are physically associated with strong couplings. It must therefore be possible to rephrase the above evolutions in terms of a strong (continuous)
coupling. 
Let 
\begin{equation}
H_K= H + K H_{\mathrm{c}} ,
 \label{eq:HKcoup}
\end{equation}
where $H$ is the Hamiltonian of the system, $H_{\mathrm{c}}$ an interaction Hamiltonian performing a "continuous
measurement" and $K$ a coupling constant.

We are interested in the $K\to\infty$ limit of the evolution operator
\begin{eqnarray}
U_{K}(t) = \e^{-\ii t H_K } 
\sim \e^{-\ii t H_{\mathrm{Z}}}  \e^{-\ii t K H_{\mathrm{c}}},
 \qquad \text{as } K\to\infty,
\label{eq:measinter}
\end{eqnarray}
where $H_{\mathrm{Z}}$ is given in Eq.~(\ref{eq:eqHz}), $P_n$ being the spectral projection of $H_{\mathrm{c}}$ (taken with a discrete spectrum)
\begin{equation}
\label{eq:diagevol}
H_{\mathrm{c}} = \sum_n \eta_n P_n, \qquad (\eta_n\neq\eta_m,
\quad \mbox{for} \; n\neq m) \ .
\end{equation}
These results are consequences of the adiabatic theorem~\cite{ref:KatoAdiabatic}.
The Zeno subspaces are again a consequence of the wildly oscillating phases between different eigenspaces.
One can interpret the above results by saying that the external field/potential takes a steady, "persuading gaze" at the system.

\section{Equivalence between continuous and pulsed formulations}
\label{sec-unitequiv}

The similarity between the "kicked" and "continuous"
formulations, outlined in Secs. \ref{sec-kickform} and \ref{sec-contform}, is in fact even more profound.
The two procedures are physically equivalent, and only differ in the order in which two limits are computed~\cite{ref:BBZeno}. 

The continuous case considers Eq.~(\ref{eq:HKcoup}) in the strong coupling limit $K\rightarrow \infty$, while the kicked dynamics is generated by
the time-dependent Hamiltonian
\begin{equation}
H_{\mathrm{kick}} =H+\tau_{0} \sum_n \delta (t - n\tau) H_{\mathrm{c}}, 
 \label{eq:puls}
\end{equation}
where $\tau$ is the period between two kicks and $U_{\mathrm{kick}}=\exp
(-\ii\tau_0 H_{\mathrm{c}})$. The $N\rightarrow \infty$ limit 
in~(\ref{eq:BBevol}) corresponds to $\tau\rightarrow 0$. The Hamiltonians~(\ref{eq:HKcoup}) and~(\ref{eq:puls}) are both limiting
cases of the following one
\begin{equation}
  H(\tau,K) = H+K
  \sum_{n}g\big( t-n(\tau+\tau_0/K) \big) H_{\mathrm{c}} ,
\label{eq:contpuls}
\end{equation}
where $g(t)=\chi_{[-\tau_0/2 K,\tau_0/2 K ]}(t)$, with a fixed $\tau_0>0$. In Eq.~(\ref{eq:contpuls}) the period between two kicks is $\tau_{0}/K+\tau$,
while the kick lasts for a time $\tau_{0}/K$.
By taking the limit $\tau\rightarrow 0$ in Eq.\ (\ref{eq:contpuls}) (a sequence of pulses of finite duration $\tau_0/K$ without any time interval among them), one recovers the continuous case (\ref{eq:HKcoup}): $H(\tau, K) \to H_K$, as $\tau\to 0$. One takes afterwards  the strong coupling limit
$K\rightarrow \infty $ and gets the Zeno dynamics and subspaces. Let us now invert the order of the limits: first take the $K\rightarrow \infty $ limit (short pulses, but with the same global---integral---effect), to obtain the kicked case~(\ref{eq:puls}): $H(\tau, K) \to H_{\mathrm{kick}}$, as $K\to\infty$. Then, take the vanishing time interval limit $\tau\rightarrow 0$ to get again the Zeno dynamics and subspaces.

In short, by denoting with $U_{\tau, K}(t)$ the unitary evolution generated by $H(\tau,K)$, the formal equivalence between the two protocols is
expressed by the relation
\begin{equation}
\label{eq:limits}
  \lim_{K\rightarrow \infty }\;\lim_{\tau \rightarrow
    0} U_{\tau,K}(t)
  =\lim_{\tau\rightarrow 0}\;\lim_{K\rightarrow \infty }U_{\tau,K}(t),
\end{equation}
with the left (right) side expressing the continuous (pulsed) case. 
This equivalence is only valid in the limit. It is physically legitimate if the inverse Zeno regime~\cite{ref:InverseZeno} is avoided.
In practice, for finite $N$ and $K$, there can be important differences \cite{ref:ControlDecoZeno}.

\section{Comments and conclusions}

The above discussion hinged upon Schr\"odinger equations, unitary dynamics, and projections \textit{\`a la} von Neumann. This is only a part of the whole picture.
Most results can be generalized to master (GKLS \cite{ref:GKLS-DariuszSaverio}) equations, quantum semigroups, and sequences of generic quantum operations~\cite{ref:ZanardiDFS-PRL2014,ref:ZanardiDFS-PRA2017,ref:unity1,ref:unity2}. Moreover, the analysis can be extended to multidimensional spaces, highlighting interesting relations with geometry \cite{ref:berry,ref:geo} and complexity \cite{ref:plato}.  

American composer Harold Budd once said: "But that's fine, because I like to have control of the ambience."  Lucky him. Physicists and mathematicians can only endeavor to control the system, out of the (often detrimental) effects of the environment.

\vspace{6pt}

\acknowledgments{We thank many colleagues with whom we discussed the quantum Zeno effect and quantum control over the last 20 years.
A special thank to our students G. Gramegna, D. Lonigro, D. Pomarico and G. Scala for their constant interest and for motivating our research.
This work is partially supported by Istituto Nazionale di Fisica Nucleare (INFN) through the project "QUANTUM". PF is partially supported by the Italian National Group of Mathematical Physics (GNFM-INdAM).}



\reftitle{References}

\end{document}